\documentclass[cameraready]{Interspeech}

\usepackage{caption}
\usepackage{subcaption}

\DeclareCaptionLabelFormat{figsub_num}{Fig.~\thefigure (#2)}
\title{Sensitivity Analysis of Generative Spatial Audio Metrics
: A Study on Responsiveness, Smoothness, and Symmetry
}

\author[affiliation={1}, orcid=0000-0003-0351-6574]{Purnima}{Kamath}
\author[affiliation={1}, orcid=0009-0003-3865-6910]{Adrian S.}{Roman}
\author[affiliation={2}, orcid=0009-0002-9964-4924]{Koichi}{Saito}
\author[affiliation={1,2,3}, orcid=0000-0002-6806-6140]{Yuki}{Mitsufuji}
\author[affiliation={1}, orcid=0000-0001-8561-5204]{Juan P.}{Bello}

\address{
    $^1$ New York University, New York, USA \\
    $^2$ Sony AI, New York, USA, $^3$ Sony Group Corporation, Tokyo, Japan
}

\email{pk3251@nyu.edu}


\usepackage{color}
\usepackage{soul}
\usepackage{subcaption}
\usepackage{multirow}
\usepackage{pifont}
\usepackage{xcolor}

\usepackage{comment}
\usepackage{float}

\newcolumntype{P}[1]{>{\centering\arraybackslash}p{#1}}

\begin{document}

\maketitle

\begin{abstract}
\vspace{-0.2em}
   Evaluating generative spatial audio for First-Order Ambisonics (FOA) remains challenging due to a limited understanding of how metrics respond to changes in spatial parameters such as azimuth and elevation. We propose a framework to analyze metric sensitivity along continuous spatial trajectories, drawing on principles of sensitivity analysis in parametric sound synthesis. Using controlled FOA scenes with increasing scene complexity, we define three desiderata for metric behavior: Responsiveness, Smoothness, and Symmetry. We assess standard distribution-based and sample-based metrics, including Fréchet Audio Distance (FAD), intensity vectors, and acoustic maps. Our findings show that FAD using localization-specific embeddings and acoustic maps yield high Responsiveness and robust Smoothness and Symmetry across conditions, while intensity vectors degrade with increasing scene complexity. This is the first step towards investigating the sensitivity of metrics for generative spatial audio.
\end{abstract}

\section{Introduction}
Spatial audio generation for First-Order Ambisonics (FOA) has recently attracted growing interest, driven by applications in immersive media and interactive machine listening~\cite{correa2023spatial,gramaccioni2024l3das23}.
The spatial and multi-channel nature of these sounds makes the generative modelling task significantly harder compared to monophonic sounds. Specifically in the spatial audio context, we expect the sources in the generated sounds to be well localized and organized in space, often in response to explicit control parameters such as azimuth and elevation~\cite{heydari2025immersediffusion,diffsage,sun2025both,kim2025visage}. While most generative models expose such controls, evaluating whether the models actually follow them remains difficult~\cite{zhu-etal-2025-asaudio}.

A key challenge is the lack of agreed‑upon metrics for spatial audio generation. Existing work has adapted distribution‑based metrics such as the Fréchet Audio Distance (FAD)~\cite{kilgour19_interspeech,gui2024fadtk} and its variants~\cite{kad}, as well as sample-based metrics, such as log spectral distance~\cite{zhang2025diffstereo} and intensity vectors~\cite{heydari2025immersediffusion,diffsage}. However, it is not well understood how sensitive or well-behaved these metrics are to changes in spatial control parameters. This lack of understanding makes it difficult to agree upon a set of metrics that accurately reflect model performance. 

Prior work on parameter sensitivity~\cite{gupta2022_paramsense} systematically varied control parameters for monophonic audio textures and examined how metrics respond. Parametric sound synthesizers~\cite{serra1990spectral} have traditionally employed sensitivity analysis to quantify the degree of control over synthesis parameters. In this framework, response curves are evaluated based on various criteria such as the magnitude of change in the output relative to unit changes in input, their monotonic nature~\cite{kamath2024sound}, and the smoothness of the curve (or the absence of jitter)~\cite{wyse2022sound}. These ideas, however, have not been applied to evaluation metrics for generative spatial audio along controlled spatial trajectories or varying scene layouts.

This work bridges these strands by proposing a meta-evaluation framework to analyze metric behavior over spatial control trajectories for generative spatial audio. We define and quantify three desiderata for parameter-sensitive spatial audio evaluation metrics: (1) \textbf{Responsiveness} of the metric to changes in the spatial parameter along a trajectory, (2) \textbf{Smoothness} based on pairwise neighbor distances, and (3) \textbf{Symmetry} between forward and reverse trajectories. Responsiveness quantifies the magnitude of metric change per unit change in input; Smoothness captures the regularity, or the absence of local discontinuities in the distance curve; and Symmetry measures how closely paired samples at opposite spatial positions match.

In addition, we advance a sensitivity study of a representative sample of metrics from the literature using a custom dataset of controlled variations of simple FOA scenes with increasing levels of complexity: single‑source sweeps, multi-source sweeps, and multiple instances of the same source. Through our analysis, we identify a few metrics that show sensitivity across all three criteria and robustness to changes in scene complexity.

\section{Method: Responsiveness, Smoothness, and Symmetry}
By sensitivity, we mean the degree to which a metric reflects changes in the signal as synthesis parameters vary sequentially. Sensitivity measures should indicate how granularly a metric distinguishes between a generated scene and a reference, with distances approaching zero as the generation matches the reference and increasing as it diverges. In spatial audio generation, this specifically concerns how accurately spatial relationships between output and reference are captured.  

We therefore expect an ideal metric to follow a ``tent-like'', unimodal progression as in Fig.~\ref{fig:expectation}: deviations from the reference spatial parameters should yield proportional, monotonic (non-abrupt) changes in distance. Moreover, the geometry of the target should be mirrored in the geometry of the output. These expectations motivate our definitions of Responsiveness, Smoothness, and Symmetry.

\noindent \textbf{Problem setup and notation}: We consider a sequence of FOA samples, $x_i,  i:[1, N]$, generated by sequentially and uniformly varying azimuth or elevation between $[-180^\circ, 180^\circ]$. For each $i^{th}$ sample, we define its distance from the $j^{th}$ sample as $d_i^j = d(x_i, x_j)$, where $d(.,.)$ is a metric under evaluation. All distances are z-score normalized per metric for a control trajectory to remove scale differences across metrics (e.g., bounded sample-based metrics versus unbounded distribution-based metrics) while preserving the shape of their variation along the control trajectory. 

\subsection{Responsiveness}
Responsiveness quantifies a metric's sensitivity to changes in azimuth or elevation. For each metric, we model the standardized distance for the $i^{th}$ sample from $x_{1}$ (the starting point of the trajectory at $-180^\circ$) defined by $d_i^{1} = d(x_i, x_{1})$ by fitting a low-order smooth function ${f}(\Delta\phi)$, where $\Delta\phi$ represents the mean angular displacement of one or more sound events in the sample $x_i$ from the reference $x_{1}$ at position $\phi_{1} = -180^\circ$ along a circular trajectory as shown in Fig.~\ref{fig:trajectory}. We model the distances computed by each metric as following the progression described in Fig.~\ref{fig:expectation}, rising from zero at the reference position, peaking near the midpoint $0^\circ$, and returning to zero after a full rotation to $180^\circ$, which we model as:
\begin{equation}
	f(\Delta\phi) = a - b * \vert \Delta\phi -c \vert,
\end{equation}
where $a$ and $b$ are estimated by the fitted curve and give the peak height of the curve and the magnitude of the slope, respectively. $c$ is the peak center and is set to the center point of the trajectory ($\Delta\phi$ at $0^\circ$). $*$ indicates scalar multiplication. The visualization in Fig.~\ref{fig:trajectory} shows the circular trajectory for both single and multi-source sweeps. We define Responsiveness as the mean of the absolute slope ($f'(\Delta\phi)$) of this fitted curve, penalized by its quality of fit (coefficient of determination) $R^2_{f}$: 
\begin{align}
f'(\Delta\phi) &= \begin{cases}
	  + b, & \Delta\phi < c\\
    - b, & \Delta\phi > c\\
    0, & \Delta\phi = c
\end{cases}\\
\textrm{Responsiveness} &=  \frac{1}{N}\sum_{i=1}^{N} \vert f'(\Delta\phi)\vert * R^2_{f} 
\end{align}
\subsection{Smoothness}
Smoothness captures the regularity of distances or the absence of jitter/sharp discontinuities along the trajectory. For this, we compute distances between neighboring samples along the trajectory. If $d^{i-1}_i$ is the normalized distance between a sample at $i$ and its neighbor at $i-1$, we compute the standard deviation of the squared distance and quantify smoothness as shown below. We square the distances so that large breaks/discontinuities are penalized more than smaller breaks.
\begin{equation}
\textrm{Smoothness} = \left(1 + \sqrt{\frac{1}{N} \sum_{i=2}^{N} \left( (d_{i}^{i-1})^2 - \frac{1}{N} \sum_{j=2}^{N} (d_{j}^{j-1})^2 \right)^2}\right)^{-1}
\end{equation}
\vspace{-2.em}
\subsection{Symmetry}
Symmetry measures how similar the metric is across the opposite left and right halves of the trajectory sweep. For each pair of samples $x_{i}$ and $x_{N-i}$ at symmetric angles $\phi_i$ and $-\phi_i$, with normalized distances $d_{i}^{1}$ and $d_{N-i}^{1}$ from $x_{1}$, we expect $d_{i}^{1} \approx d_{N-i}^{1}$. We define Symmetry based on the RMSE of the distances as follows:
\begin{align}
    \textrm{Symmetry Error } (SE) &= \frac{\sqrt{\frac{2}{N}\sum_{i=1}^{N/2}((d_{i}^{1}-d_{N-i}^{1})^2)}}{\frac{2}{N}\sum_{i=1}^{N/2}(d_{i}^{1}-d_{N-i}^{1})}\\
    \textrm{Symmetry} &= e^{-SE}
\end{align}
\noindent The inverse exponential of $SE$ binds the score between $[0,1]$, especially for metrics that may exhibit very high asymmetry. Higher values are better for all measures.

\section{Experimental Design}
\begin{figure}[t]
    \centering
    \begin{subfigure}{0.35\textwidth}
        \centering
     \includegraphics[width=0.7\linewidth]{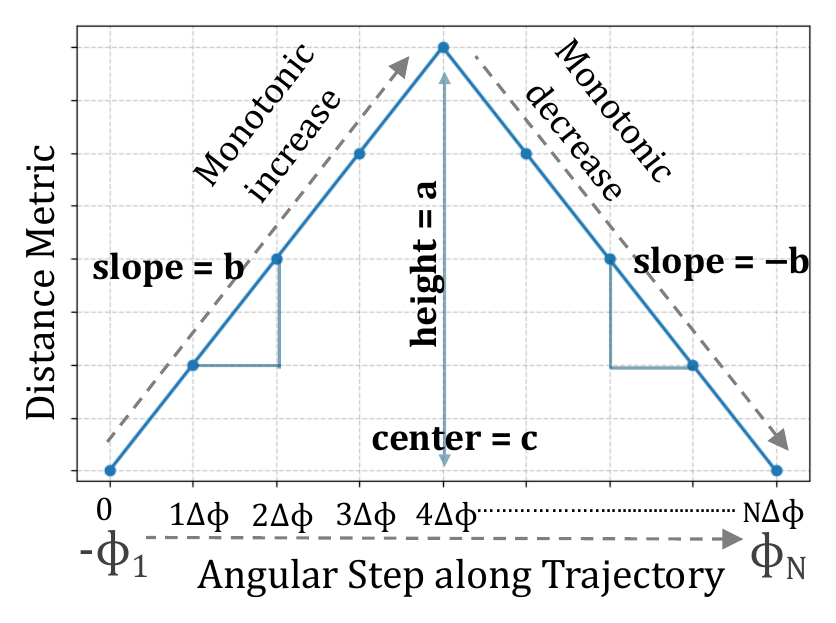}
     \caption{Expectation of the Response Curve.
     }\label{fig:expectation}
    \end{subfigure}
    \begin{subfigure}{0.45\textwidth}
    \centering
     \includegraphics[width=0.9\linewidth]{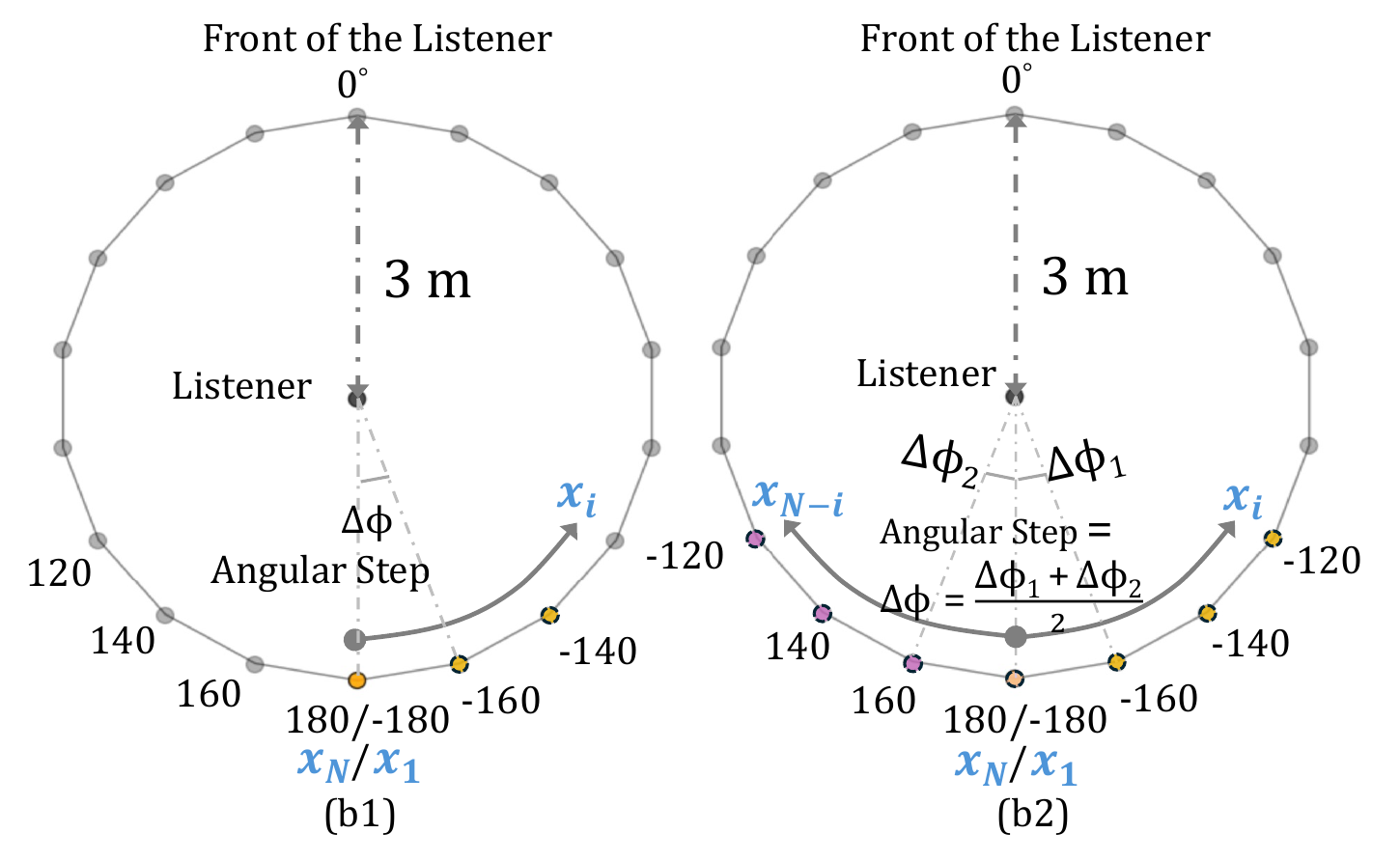}
     \caption{Trajectory for varying azimuth for (b1) single source, and (b2) multiple-source events experiments. Elevation trajectories follow similar sweep patterns.
     }\label{fig:trajectory}
 \vspace{-1.5em}
    \end{subfigure}
\end{figure}
We conduct experiments to understand two things: (1) the sensitivity of the metrics as spatial parameters vary along a control trajectory, and (2) their robustness to increasing scene complexity and noise. To this end, we create a large set of precisely controlled synthetic scene variations, deploy a representative set of metrics across them, and measure the sensitivity of these metrics to the controlled variables. 

\begin{figure*}[t]
\centering
 \includegraphics[width=0.95\linewidth]{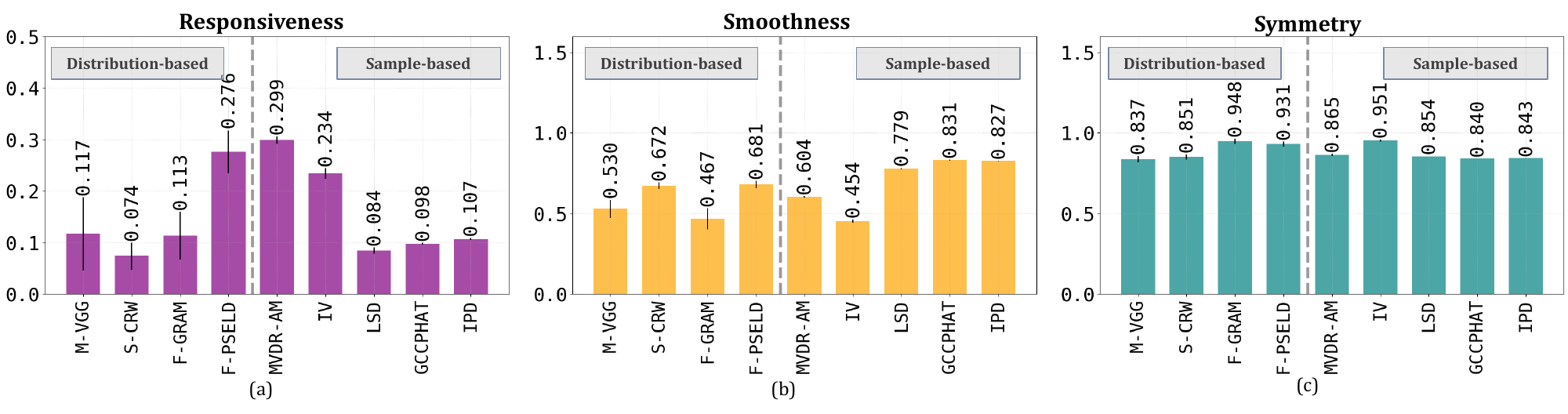}
 \vspace{-0.9em}
 \caption{Results across all experimental conditions. Higher values are better. Standard error bars computed by bootstrapping.
 }\label{fig:sec1_allexp}
 \vspace{-1.5em}
\end{figure*}

\subsection{Data Generation}
We use FOA Room Impulse Responses (RIRs) from SoundSpaces 1.0~\cite{chen2020soundspaces} and spatialize sounds using SpatialScaper~\cite{roman2024spatial}. SoundSpaces contains pre-simulated RIRs for different source-listener positions in Matterport 3D scenes, with a grid resolution of \SI{1}{meter} at a \SI{16}{kHz} sampling rate. We select $30$ largest scenes, position the listener at the center of the densest part of the scene, and move the source along circular azimuth or elevation trajectories at a $3$ m radius, as shown in Fig.~\ref{fig:trajectory}. Azimuth and elevation are linearly interpolated in $[-180^\circ, 180^\circ]$ with a step size of $20^\circ$ (total $19$ steps). When sweeping azimuth, elevation is fixed to the horizontal plane; when sweeping elevation, azimuth is fixed at $0^\circ$. Given the grid resolution of our RIR dataset and radius of \SI{3}{meters}, our step size $\Delta\phi = 20^\circ$ ensures each RIR source position is selected only once along the sweep.

For monophonic sound events, we curate single-source stems from FSD50K~\cite{fonseca2021fsd50k} (as in~\cite{hu2025pseldnets}). We segment clips into 10-second excerpts and increase sound event density in each sample by repeating short events along the time axis. We randomly select 30 classes from this set to synthesize our sounds.

We synthesize sounds for six experimental conditions by varying azimuth and elevation under three spatial layouts of increasing polyphony and scene complexity:

\begin{itemize}
\item \textbf{Single Source (SS)}: This experiment isolates how each metric responds to a single moving source around a listener. We randomly select a monophonic sound event and convolve it with RIRs using SpatialScaper, varying the spatial parameter along a trajectory from $[-180^\circ, 180^\circ]$.

\item \textbf{Multiple Sources (MS)}: Two events from different classes follow a counter-rotating trajectory as shown in Fig.~\ref{fig:trajectory}. One source sweeps $-180^\circ \rightarrow 180^\circ$, the other $180^\circ \rightarrow -180^\circ$, yielding co-located or symmetrically opposed sources of different classes at each step.  

\item \textbf{Single Source - Multiple Instances (SSMI)}: This experiment stresses the metrics under symmetric multi-instance trajectories of the same class that spatially mirror each other. Two events from the same class follow the same counter-rotating trajectories as in the MS condition. 

\end{itemize}

For each of the three setups, we repeat the experiment with added noise (\textbf{SS+N}, \textbf{MS+N}, and \textbf{SSMI+N}) and evaluate all sensitivity measures\footnote{\url{https://github.com/pkamath2/sa\_sensitivity}} on the noisy signals. Signals are generated with a sampling rate of \SI{16}{kHz}. For noise experiments, we add Gaussian noise with a random SNR between $0$ and \SI{15}{dB}. Overall, we synthesize scenes for   
$3$ experiments × $2$ variations (azimuth and elevation) × $30$ classes × $10$ clips per class × $19$ trajectory steps × $2$ conditions (clean and noisy), for a total of $68,400$ ten-second FOA samples.

\subsection{Metrics Under Assessment}
Currently, a wide variety of metrics have been used to evaluate generative spatial audio models. In our experiments, we focus on open-source (and weights) metrics most commonly used in prior work, complemented by a small set of additional metrics. We assess two families of metrics in our sensitivity analysis.

\noindent\textbf{Distribution-based Metrics}: We use FAD to compute distances between the embeddings of FOA scenes at a reference position and target positions along spatial trajectories. We compute metrics by extracting embeddings from four networks: M-VGG based on VGGish~\cite{vggish_hershey2017cnn}, S-CRW from StereoCRW~\cite{stereoCRW_chen2022sound}, F-GRAM from GRAM~\cite{yuksel2026gram}, and F-PSELD from PSELDNets~\cite{hu2025pseldnets}, across mono, stereo, FOA formats. VGGish embeddings were obtained by averaging across FOA channels to cancel directionality and produce monophonic sounds, while StereoCRW embeddings were generated via FOA-to-stereo conversion (L=W+Y, R=W–Y) leveraging implementation from~\cite{saito2025soundreactor}. GRAM provides self-supervised FOA representations optimized for input reconstruction, and PSELDNets extend Hierarchical Token-Semantic Audio Transformer (HTS-AT)~\cite{chen2022hts} to predict multi‑ACCDOA~\cite{multiaccDOA_shimada2022} targets. VGGish and StereoCRW use log‑mel spectrogram inputs, whereas GRAM and PSELDNets combine log‑mel spectrograms with intensity vectors. We also evaluated KAD~\cite{kad} on those embeddings but observed negative values and numerical instability under noise. We leave its applicability for sensitivity analysis to future work.
\\
\noindent\textbf{Sample-based Metrics}: We evaluate a range of phase, magnitude, and spatial acoustic maps-based metrics, including Interchannel Phase Differences (IPD), Log Spectral Distances (LSD), Intensity Vectors (IVs), and Generalized Cross-Correlation Phase Transform (GCCPHAT). We compute L2 distances over these metrics. Additionally, we use MVDR-AM (minimum-variance distortionless response) beamforming-based acoustic maps from~\cite{mccormack2017parametric}. As in SPARTA~\cite{mccormack2019sparta}, we use the MVDR-AM as a means to provide a 2D time-averaged spatial distribution of events in a scene, and thus compute LPIPS~\cite{zhang2018unreasonable} (perceptual distances) over them. 

\begin{table}[H]
  \scriptsize
  \caption{All metrics at a glance}
  \vspace{-0.9em}
  \label{tab:responsiveness}
  \centering
  \begin{tabular}{P{5.5em}P{5.5em}P{9em}P{4.5em}}
    \toprule
    Metric&Format&Embedding&Distance\\
    \midrule
    \textbf{M-VGG}&\textbf{M}ono (1-ch)&\textbf{VGG}ish~&FAD\\
    \textbf{S-CRW}&\textbf{S}tereo (2-ch)&Stereo\textbf{CRW}&FAD\\
    \textbf{F-GRAM}&\textbf{F}OA (4-ch)&\textbf{GRAM}&FAD\\
    \textbf{F-PSELD}&\textbf{F}OA (4-ch)&\textbf{PSELD}Nets&FAD\\
    \midrule
    \textbf{MVDR-AM}&FOA (4-ch)&\scriptsize{2D \textbf{A}coustic \textbf{M}aps}&LPIPS\\
    \textbf{IV}&FOA (4-ch)&-&L2\\
    \textbf{GCCPHAT}&FOA (4-ch)&-&L2\\
    \textbf{LSD}&FOA (4-ch)&-&L2\\
    \textbf{IPD}&FOA (4-ch)&-&L2\\
    \bottomrule
  \end{tabular}
\end{table}

\section{Results \& Discussion}
\noindent \textbf{Main Comparisons:} Fig.~\ref{fig:sec1_allexp} (a–c) summarizes Responsiveness, Smoothness, and Symmetry across all experimental conditions. Each bar plot shows the mean scores across azimuth and elevation sweeps, averaged across all conditions. 

For sample‑based metrics, the Responsiveness plot shows that MVDR‑AM achieves the highest scores, followed by IV, whereas LSD, GCCPHAT, and IPD exhibit consistently low Responsiveness. Although the poor performance of LSD is unsurprising, given that magnitude spectrograms carry no explicit spatial information, the low scores of GCCPHAT and IPD, despite their spatial nature and presumed sensitivity to spatial variation, appear to stem from their susceptibility to noise. 

For distribution‑based metrics, spatially informed F‑PSELD attains the highest mean Responsiveness, outperforming metrics with little or no explicit spatial structure, such as M‑VGG and S‑CRW. Although F‑GRAM is derived from a canonical FOA GRAM representation, it attains comparatively low scores in our setting. We view this primarily as an artefact of our evaluation setup, which deliberately stress-tests metrics on artificial scenes and noise conditions.
Within this setting, localization‑driven training objectives (as used in F-PSELD) appear to yield more robust metrics than GRAM’s reconstruction‑based objective.

For Symmetry, most distribution‑based and sample‑based metrics achieve strong scores, indicating that their forward and reverse trajectories are largely consistent and suggesting that Symmetry alone is an insufficient indicator of sensitivity and may misrepresent other aspects of control behavior.

\begin{figure}[t]
\centering
 \includegraphics[width=0.99\linewidth]{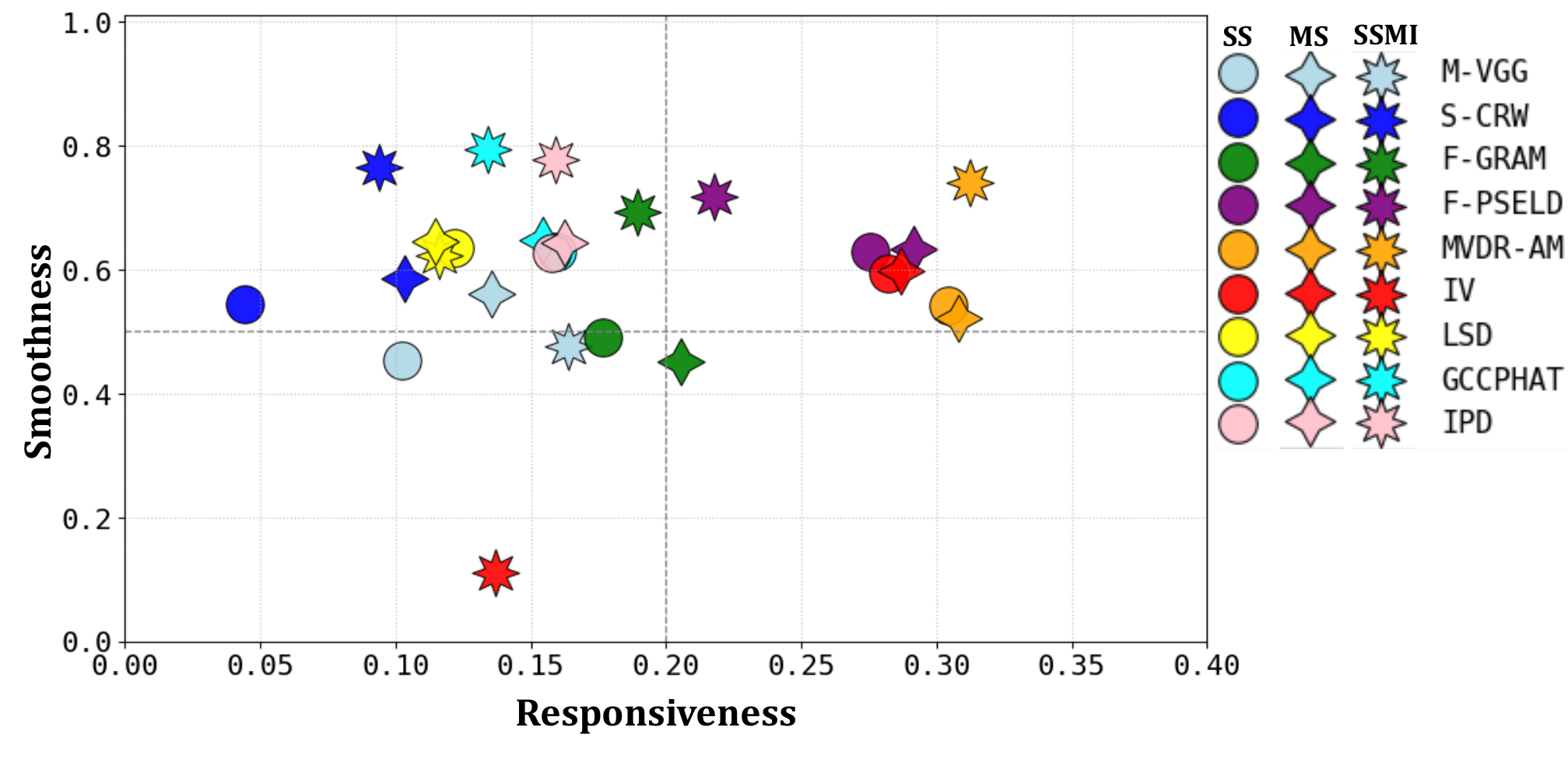}
 \caption{Responsiveness vs. Smoothness Trade-off in clean conditions. The right upper quadrant indicates high scores.
 }\label{fig:sec1_scatterplot}
 \vspace{-1.6em}
\end{figure}

\noindent\textbf{Responsiveness vs. Smoothness Trade-off}: For Smoothness, in Fig.~\ref{fig:sec1_allexp}(b), amongst the metrics that exhibit high Responsiveness, namely F-PSELD, MVDR-AM, and IV, only moderate Smoothness scores are observed. In contrast, metrics such as GCCPHAT, LSD, and IPD achieve the highest Smoothness scores but very low Responsiveness scores. To examine this trade‑off, Fig.~\ref{fig:sec1_scatterplot} plots Responsiveness vs. Smoothness under clean conditions. Metrics in the upper‑right quadrant (high Responsiveness and Smoothness) are particularly desirable: MVDR‑AM and F‑PSELD consistently occupy this region across scene complexities, and IV does so except in the SSMI condition, where it shifts toward lower Responsiveness and Smoothness. For distribution‑based metrics, F‑PSELD and F‑GRAM exhibit higher Responsiveness than M‑VGG and S‑CRW, indicating unsurprisingly that, under clean conditions, training on 4‑channel data more effectively captures spatial variation along the control trajectory than mono or stereo data.

\begin{figure}[t]
\centering
 \includegraphics[width=1.0\linewidth]{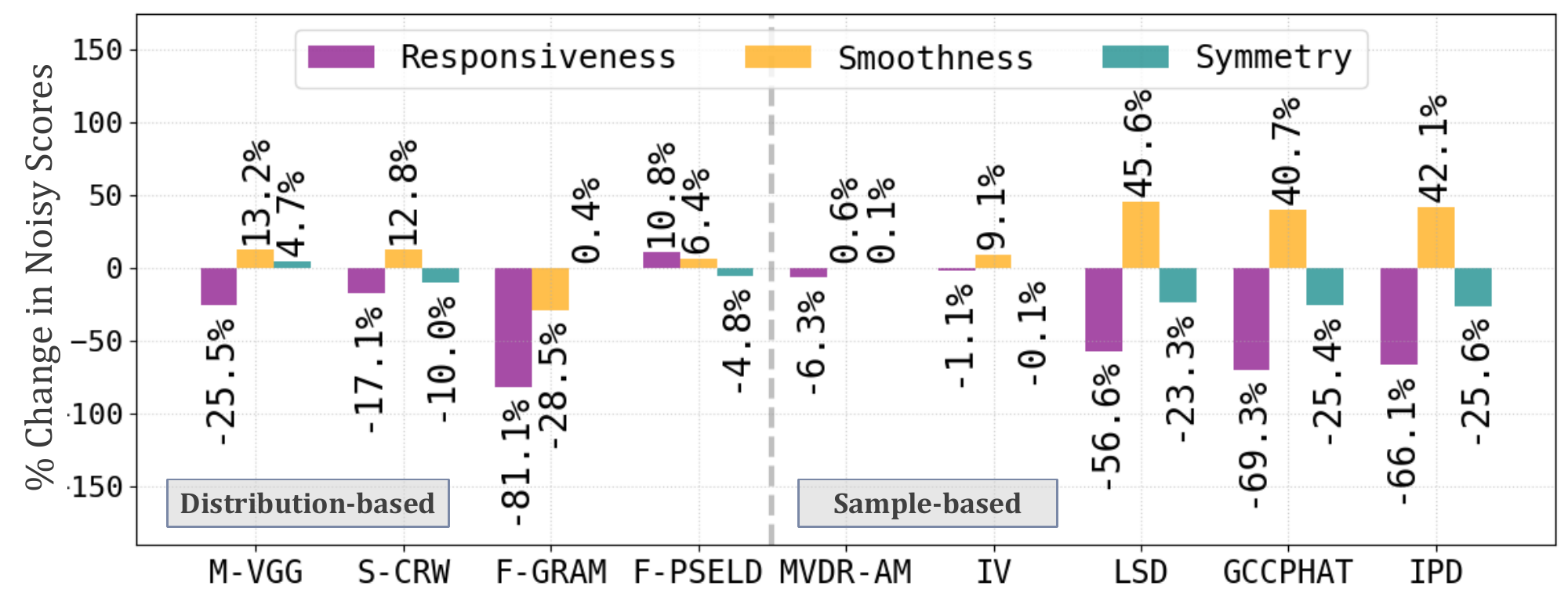}
 \caption{\% Change in Scores w/ Additive Noise. Changes in scores closer to 0\% indicate greater robustness in the metrics.
 }\label{fig:sec3_robustness_noise}
 \vspace{-0.1em}
\end{figure}

\noindent\textbf{Robustness to Noise}: We examine how additive noise along control trajectories affects metric sensitivity. For each metric, we compute the percentage change in sensitivity scores under noisy conditions relative to clean conditions (Fig.~\ref{fig:sec3_robustness_noise}). Among sample-based metrics, MVDR‑AM and IVs exhibit the ideal behavior of minimal score change, indicating strong noise robustness. For distribution-based metrics, F‑PSELD exhibits the lowest average change, likely because its embeddings were trained on both IVs and spectrograms, which enhance spatial and spectral stability. F‑PSELD also outperforms F‑GRAM, indicating that PSELDNets were more robust to diffuse-noise perturbations leveraged in our evaluation. In contrast, sample-based metrics such as LSD, GCCPHAT, and IPD exhibit reduced Responsiveness, increased Smoothness, and lower Symmetry, indicating that their response curves are mostly flat and smooth under noise, obscuring meaningful control behavior.

\begin{figure}[t]
\centering
 \includegraphics[width=0.99\linewidth]{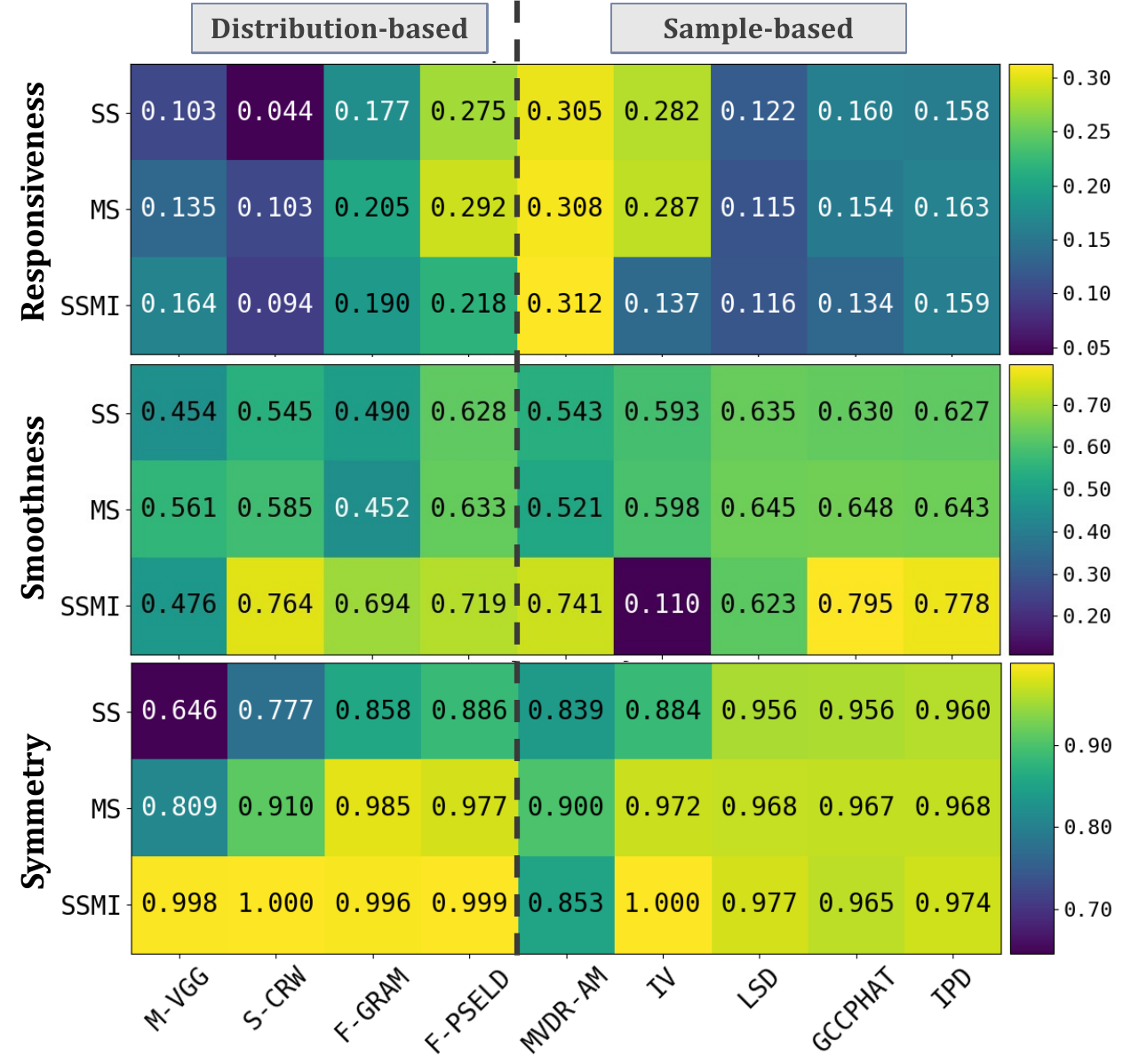}
 \caption{Robustness to Source Complexity in clean conditions. 
 }\label{fig:sec4_heatmaps}
 \vspace{-1.7em}
\end{figure}

\noindent\textbf{Robustness to Source Complexity}: We study how source complexity affects metric sensitivity under clean conditions. Fig.~\ref{fig:sec4_heatmaps} shows Responsiveness, Smoothness, and Symmetry scores ordered by increasing scene complexity. With SS as reference, MVDR‑AM exhibits the strongest and most stable Responsiveness, while F‑PSELD and IVs remain stable from SS to MS but degrade under SSMI. For Smoothness, all metrics are stable across SS and MS, but SSMI causes IV distance curves to collapse and introduce large discontinuities, lowering both Responsiveness and Smoothness while increasing Symmetry. This suggests IVs are highly sensitive to mirrored‑source cancellations and may not be suitable for specific cases involving symmetric multi-source evaluations. In contrast, Smoothness for F‑PSELD and F‑GRAM (both trained on IVs alongside log‑mel spectrograms) remains stable, indicating that their combined use of IVs and log‑mel spectrograms helps mitigate IV collapse in the learned representations.

\section{Conclusion}
\vspace{-0.2em}
In this work, we defined sensitivity as the Responsiveness, Smoothness, and Symmetry of evaluation metrics under controlled spatial parameter changes and conducted an empirical study of their behavior. Localization-based metrics such as F‑PSELD, IV, and MVDR‑AM showed strong Responsiveness with good Smoothness trade-off, and were robust to noise and scene complexity, although IV degraded under symmetric multi-source evaluations. For distribution-based metrics, higher spatial input resolution, localization-oriented training objectives (such as SELD-based), and embeddings from networks trained on IVs together with spectrograms improved sensitivity, while sample-based magnitude or phase metrics (LSD, IPD) and noise-sensitive measures (GCCPHAT) were less reliable. This study is limited to artificially synthesized FOA data and a small set of metrics; future work will extend the framework to a broader set of metrics, denser RIR sampling, real-world data, room geometry, and perceptual validation. This study constitutes a first step towards understanding the sensitivity of metrics for evaluating generative spatial audio models.

\section{Acknowledgments}
This work is partially funded by the NYU / SONY Audio Institute for Music Business and Technology.

\section{Use of Generative AI Disclosure}
In preparing this work, the authors used Claude Code and Perplexity AI as tools for literature exploration, sentence paraphrasing, and drafting code, after which they carefully reviewed and revised the content before using it within their framework and manuscript. The authors accept full responsibility for the content in this publication.


\bibliographystyle{IEEEtran}
\bibliography{paper}

@inproceedings{kilgour19_interspeech,
  title     = {{Fréchet Audio Distance: A Reference-Free Metric for Evaluating Music Enhancement Algorithms}},
  author    = {Kevin Kilgour and Mauricio Zuluaga and Dominik Roblek and Matthew Sharifi},
  year      = {2019},
  booktitle = {{Interspeech}},
  pages     = {2350--2354},
  doi       = {10.21437/Interspeech.2019-2219},
  issn      = {2958-1796},
}

@inproceedings{gui2024fadtk,
  title={{Adapting Fréchet Audio Distance for generative music evaluation}},
  author={Gui, Azalea and Gamper, Hannes and Braun, Sebastian and Emmanouilidou, Dimitra},
  booktitle={ICASSP 2024-2024 IEEE International Conference on Acoustics, Speech and Signal Processing (ICASSP)},
  pages={1331--1335},
  year={2024},
  organization={IEEE}
}

@article{yuksel2026gram,
  title={{GRAM}: Spatial general-purpose audio representations for real-world environments},
  author={Yuksel, Goksenin and van Gerven, Marcel and van der Heijden, Kiki},
  journal={arXiv preprint arXiv:2602.03307},
  year={2026}
}

@ARTICLE{hu2025pseldnets,
  author={Hu, Jinbo and Cao, Yin and Wu, Ming and Kang, Fang and Yang, Feiran and Wang, Wenwu and Plumbley, Mark D. and Yang, Jun},
  journal={IEEE Transactions on Audio, Speech and Language Processing}, 
  title={{PSELDNets}: Pre-Trained Neural Networks on a Large-Scale Synthetic Dataset for Sound Event Localization and Detection}, 
  year={2025},
  volume={33},
  number={},
  pages={2845-2860},
  doi={10.1109/TASLPRO.2025.3587446}}

@inproceedings{mccormack2019sparta,
  title={{SPARTA \& COMPASS: Real-time implementations of linear and parametric spatial audio reproduction and processing methods}},
  author={McCormack, Leo and Politis, Archontis},
  booktitle={Audio Engineering Society Conference: 2019 AES International Conference on Immersive and Interactive Audio},
  year={2019},
  organization={Audio Engineering Society}
}

@inproceedings{gupta2022_paramsense,
  author       = {Chitralekha Gupta and
                  Yize Wei and
                  Zequn Gong and
                  Purnima Kamath and
                  Zhuoyao Li and
                  Lonce Wyse},
  title        = {Parameter Sensitivity of Deep-Feature based Evaluation Metrics for
                  Audio Textures},
  booktitle    = {Proceedings of the 23rd International Society for Music Information
                  Retrieval Conference, {ISMIR} 2022},
  pages        = {462--468},
  year         = {2022}
}

@inproceedings{chen2020soundspaces,
	title = {SoundSpaces: Audio-Visual Navigation in 3D Environments},
	author = {Chen, Changan and Jain, Unnat and Schissler, Carl and Gari, Sebastia Vicenc Amengual and Al-Halah, Ziad and Ithapu, Vamsi Krishna and Robinson, Philip and Grauman, Kristen},
	year = {2020},
    booktitle={European Conference on Computer Vision ECCV},
}

@inproceedings{vggish_hershey2017cnn,
  title={{CNN} architectures for large-scale audio classification},
  author={Hershey, Shawn and Chaudhuri, Sourish and Ellis, Daniel PW and Gemmeke, Jort F and Jansen, Aren and Moore, R Channing and Plakal, Manoj and Platt, Devin and Saurous, Rif A and Seybold, Bryan and others},
  booktitle={2017 {IEEE} International Conference on Acoustics, Speech and Signal Processing {ICASSP}},
  pages={131--135},
  year={2017},
  organization={IEEE}
}

@article{saito2025soundreactor,
  title={{SoundReactor}: {Frame-level Online Video-to-Audio Generation}},
  author={Saito, Koichi and Tanke, Julian and Simon, Christian and Ishii, Masato and Shimada, Kazuki and Novack, Zachary and Zhong, Zhi and Hayakawa, Akio and Shibuya, Takashi and Mitsufuji, Yuki},
  journal={arXiv preprint arXiv:2510.02110},
  year={2025}
}

@inproceedings{stereoCRW_chen2022sound,
  title={Sound localization by self-supervised time delay estimation},
  author={Chen, Ziyang and Fouhey, David F and Owens, Andrew},
  booktitle={European Conference on Computer Vision (ECCV)},
  pages={489--508},
  year={2022}
}

@INPROCEEDINGS{multiaccDOA_shimada2022,
  author={Shimada, Kazuki and Koyama, Yuichiro and Takahashi, Shusuke and Takahashi, Naoya and Tsunoo, Emiru and Mitsufuji, Yuki},
  booktitle={ICASSP 2022 - 2022 IEEE International Conference on Acoustics, Speech and Signal Processing (ICASSP)}, 
  title={{Multi-ACCDOA: Localizing And Detecting Overlapping Sounds From The Same Class With Auxiliary Duplicating Permutation Invariant Training}}, 
  year={2022},
  volume={},
  number={},
  pages={316-320},
}

@inproceedings{heydari2025immersediffusion,
  title={Immersediffusion: A generative spatial audio latent diffusion model},
  author={Heydari, Mojtaba and Souden, Mehrez and Conejo, Bruno and Atkins, Joshua},
  booktitle={ICASSP 2025-2025 IEEE International Conference on Acoustics, Speech and Signal Processing (ICASSP)},
  pages={1--5},
  year={2025},
  organization={IEEE}
}

@inproceedings{
sun2025both,
title={{Both Ears Wide Open: Towards Language-Driven Spatial Audio Generation}},
author={Peiwen Sun and Sitong Cheng and Xiangtai Li and Zhen Ye and Huadai Liu and Honggang Zhang and Wei Xue and Yike Guo},
booktitle={The Thirteenth International Conference on Learning Representations},
year={2025}
}

@inproceedings{
kim2025visage,
title={Vi{SAG}e: {Video-to-Spatial Audio Generation}},
author={Jaeyeon Kim and Heeseung Yun and Gunhee Kim},
booktitle={The Thirteenth International Conference on Learning Representations},
year={2025}
}

@inproceedings{zhang2018unreasonable,
  title={The unreasonable effectiveness of deep features as a perceptual metric},
  author={Zhang, Richard and Isola, Phillip and Efros, Alexei A and Shechtman, Eli and Wang, Oliver},
  booktitle={Proceedings of the IEEE conference on computer vision and pattern recognition},
  pages={586--595},
  year={2018}
}

@article{fonseca2021fsd50k,
  title={{FSD50k}: an open dataset of human-labeled sound events},
  author={Fonseca, Eduardo and Favory, Xavier and Pons, Jordi and Font, Frederic and Serra, Xavier},
  journal={IEEE/ACM Transactions on Audio, Speech, and Language Processing},
  volume={30},
  pages={829--852},
  year={2021},
  publisher={IEEE}
}

@inproceedings{roman2024spatial,
  title={Spatial scaper: a library to simulate and augment soundscapes for sound event localization and detection in realistic rooms},
  author={Roman, Iran R and Ick, Christopher and Ding, Sivan and Roman, Adrian S and McFee, Brian and Bello, Juan P},
  booktitle={ICASSP 2024-2024 IEEE International Conference on Acoustics, Speech and Signal Processing (ICASSP)},
  pages={1221--1225},
  year={2024},
  organization={IEEE}
}

@article{kad,
    author={Chung, Yoonjin and Eu, Pilsun and Lee, Junwon and Choi, Keunwoo and Nam, Juhan and Chon, Ben Sangbae},
    title={{KAD: No More FAD! An Effective and Efficient Evaluation Metric for Audio Generation}}, 
    journal = {arXiv:2502.15602},
    year = {2025}
}

@inproceedings{diffsage,
  author       = {Saksham Singh Kushwaha and
                  Jianbo Ma and
                  Mark R. P. Thomas and
                  Yapeng Tian and
                  Avery Bruni},
  title        = {{Diff-SAGe: End-to-End Spatial Audio Generation Using Diffusion Models}},
  booktitle    = {2025 {IEEE} International Conference on Acoustics, Speech and Signal
                  Processing, {ICASSP} 2025},
  pages        = {1--5},
  publisher    = {{IEEE}},
  year         = {2025}
}

@inproceedings{mccormack2017parametric,
  title={Parametric acoustic camera for real-time sound capture, analysis and tracking},
  author={McCormack, Leo and Delikaris-Manias, Symeon and Pulkki, Ville},
  booktitle={Proceedings of the 20th International Conference on Digital Audio Effects (DAFx-17)},
  pages={412--419},
  year={2017}
}

@inproceedings{chen2022hts,
  title={{HTS-AT: A hierarchical token-semantic audio transformer for sound classification and detection}},
  author={Chen, Ke and Du, Xingjian and Zhu, Bilei and Ma, Zejun and Berg-Kirkpatrick, Taylor and Dubnov, Shlomo},
  booktitle={ICASSP 2022-2022 IEEE International Conference on Acoustics, Speech and Signal Processing (ICASSP)},
  pages={646--650},
  year={2022},
  organization={IEEE}
}

@inproceedings{zhu-etal-2025-asaudio,
    title = "{ASA}udio: A Survey of Advanced Spatial Audio Research",
    author = "Zhu, Zhiyuan  and
      Zhang, Yu  and
      Guo, Wenxiang  and
      Pan, Changhao  and
      Zhao, Zhou",
    booktitle = "Proceedings of the 14th International Joint Conference on Natural Language Processing and the 4th Conference of the Asia-Pacific Chapter of the Association for Computational Linguistics",
    month = dec,
    year = "2025",
    pages = "417--442",
}

@inproceedings{zhang2025diffstereo,
  title={DiffStereo: End-to-End Mono-to-Stereo Audio Generation with Diffusion Transformer},
  author={Zhang, Suqi and Dai, Zheqi and Zang, Yongyi and Cao, Yin and Kong, Qiuqiang},
  booktitle={Proc. Interspeech 2025},
  pages={3150--3154},
  year={2025}
}

@article{serra1990spectral,
  title={Spectral modeling synthesis: A sound analysis/synthesis system based on a deterministic plus stochastic decomposition},
  author={Serra, Xavier and Smith, Julius},
  journal={Computer Music Journal},
  volume={14},
  number={4},
  pages={12--24},
  year={1990},
  publisher={JSTOR}
}

@inproceedings{wyse2022sound,
  title={Sound model factory: An integrated system architecture for generative audio modelling},
  author={Wyse, Lonce and Kamath, Purnima and Gupta, Chitralekha},
  booktitle={International Conference on Computational Intelligence in Music, Sound, Art and Design (Part of EvoStar)},
  pages={308--322},
  year={2022},
  organization={Springer}
}

@inproceedings{kamath2024sound,
  title={{Sound Designer-Generative AI Interactions: Towards Designing Creative Support Tools for Professional Sound Designers}},
  author={Kamath, Purnima and Morreale, Fabio and Bagaskara, Priambudi Lintang and Wei, Yize and Nanayakkara, Suranga},
  booktitle={Proceedings of the 2024 CHI Conference on Human Factors in Computing Systems},
  articleno = {730},
  numpages = {17},
  year={2024}
}

@inproceedings{correa2023spatial,
  title={{Spatial audio in virtual reality: a systematic review}},
  author={Corr{\^e}a De Almeida, Gustavo and Costa de Souza, Vinicius and Da Silveira J{\'u}nior, Luiz Gonzaga and Veronez, Maur{\'\i}cio Roberto},
  booktitle={Proceedings of the 25th symposium on virtual and augmented reality},
  pages={264--268},
  year={2023}
}

@article{gramaccioni2024l3das23,
  title={{L3das23: Learning 3d audio sources for audio-visual extended reality}},
  author={Gramaccioni, Riccardo F and Marinoni, Christian and Chen, Changan and Uncini, Aurelio and Comminiello, Danilo},
  journal={IEEE Open Journal of Signal Processing},
  volume={5},
  pages={632--640},
  year={2024},
  publisher={IEEE}
}

\end{document}